\documentclass[prl,superscriptaddress,showpacs,twocolumn]{revtex4}
\usepackage{graphicx}
\usepackage{amsmath}
\usepackage{amsfonts}
\usepackage{amssymb}
\begin{document}
\author{M. De Raychaudhury}
\affiliation{Max-Planck-Institut f\"{u}r Festk\"{o}rperforschung, Heisenbergstrasse 1,
D-70569 Stuttgart, Germany}
\affiliation{S.N. Bose National Centre for Basic Sciences, Kolkata 700098, India}
\author{E. Pavarini}
\affiliation{Institut f\"{u}r Festk\"{o}rperforschung, Forschungzentrum JŸlich, D-52425
Juelich, Germany}
\affiliation{CNISM-Dipartimento di Fisica ``A. Volta'', Universit\`a di Pavia, Via Bassi
6, I-27100 Pavia, Italy}
\author{O.K. Andersen}
\affiliation{Max-Planck-Institut f\"{u}r Festk\"{o}rperforschung, Heisenbergstrasse 1,
D-70569 Stuttgart, Germany}
\title{Orbital fluctuations  in the different phases of LaVO$_3$ and YVO$_3$}
\begin{abstract}
We investigate the importance of quantum orbital fluctuations  in the 
orthorhombic and monoclinic phases of the Mott insulators  LaVO$_3$ and YVO$_3$.
First, we construct {\it ab-initio} material-specific   $t_{2g}$ Hubbard models.
Then, by using dynamical mean-field theory, we  calculate the spectral matrix 
as a function of temperature. Our Hubbard bands and Mott gaps are in very 
good agreement with spectroscopy.
We show that in orthorhombic LaVO$_3$, quantum orbital 
fluctuations are strong and that they are suppressed {\em only} 
in the monoclinic 140$\,$K phase. In YVO$_3$  
the suppression  happens already at 300$\,$K. 
We show that Jahn-Teller {\em and}   GdFeO$_{3}$-type distortions are both crucial in determining 
the type of orbital and magnetic order in the low temperature phases.  
\end{abstract}
\pacs{71.27.+a, 71.30.+h, 71.15.Ap}
\maketitle
\begin{figure}[th]
\begin{center}
\rotatebox{0}{\includegraphics[width=0.4\textwidth]{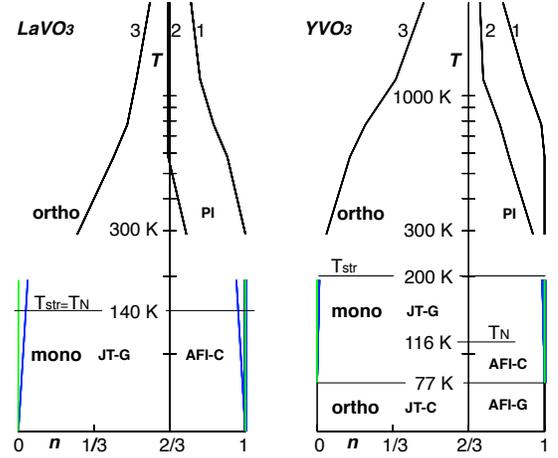}}
\end{center} 
\caption{Temperature-dependent structural and magnetic 
phases of LaVO$_{3}$ and YVO$_{3}$. The lines show 
LDA+DMFT (quantum Monte Carlo \cite{hirsch})
 results  
for the occupations,
 $n$, of the   three $t_{2g}$ crystal-field orbitals, 1, 2, and 3 (Table~\ref{cef}).  
Black lines: orthorhombic phases. Green and blue lines: 
monoclinic, sites 1 and 3 (see Fig.~\ref{orbs}).  
For each structure  we calculated the occupations down the temperature at which the orbital polarizations are essentially complete  ($T\sim 200~$K) and then extrapolated in a standard way \cite{hirsch} to T=0 K. 
}
\label{FigT} 
\end{figure}

The Mott insulating $t_{2g}^{2}$ perovskites LaVO$_{3}$ and YVO$_{3}$ exhibit an
unusual series of structural and magnetic phase transitions (Fig.~\ref{FigT}) 
with temperature-induced magnetization reversal phenomena \cite{ren2}
and other exotic properties  \cite{ulrich03,yan}.
While it is now
recognized that the V-$\,t_{2g}$  orbital degrees of freedom and the strong Coulomb
repulsion are the key ingredients, it is still controversial whether
classical (orbital order) \cite{ren2,sawada,Noguchi,Fang,solovyev06,Imada}
or quantum (orbital fluctuations) \cite{ulrich03,Khalbig} effects are responsible for
the rich physics of these vanadates.

\begin{figure}[th]
\begin{center}
\rotatebox{0}{\includegraphics[width=0.45\textwidth]{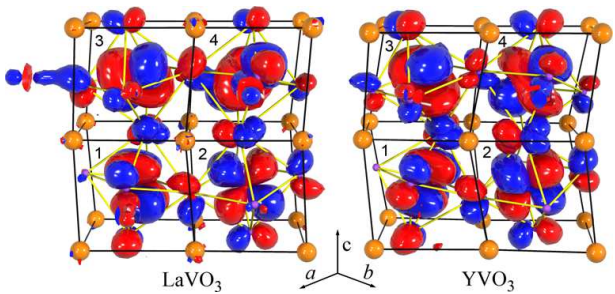}}
\end{center}
\caption{
Primitive cell containing four ABO$_{3}$ units.
A-ions are orange, B-ions are at the  centers of yellow 
O-octahedra. In terms of the primitive translation vectors, 
$\mathbf{a},$ $\mathbf{b},$ and $\mathbf{c}$ 
the global $\mathbf{x},$ $\mathbf{y}$, and $\mathbf{z}$ axes, directed
approximately along the BO bonds, are: $\mathbf{x}=\mathbf{a}/(2+2\protect
\alpha )+\mathbf{b}/(2+2\protect\beta )$ 
$\mathbf{y}\!=\!-\mathbf{a}/(2+2\protect\alpha )\!+\!\mathbf{b}/(2+2\protect\beta )$
and $\mathbf{z}=\!(\mathbf{c} \!-\mathbf{c}\cdot {\mathbf{b
}})/(2+2\protect\gamma )$.
Here, $\protect\alpha $, 
$\protect\beta $, and $\protect\gamma $ are small and $|\mathbf{x}| = |\mathbf{y
}| =|\mathbf{z}| =$3.92\AA\ (3.82\AA ) for LaVO$_{3}$ (YVO$_{3}$).
 The B-containing $bc$-plane glide-mirrors  (with translation $\left( \mathbf{b}\!-\!
\mathbf{a}\right) /2$) unit 1 in 2 and 3 in
4, and exchanges the local, B-centered $x$ and $y$ coordinates. In the
orthorhombic Pbnm structure (but not in the monoclinic P2$_{1}$/a structure), the A-containing $ab$-plane  mirrors  unit 1 in 3 and 2 in 4. 
The empty crystal-field orbitals $\left\vert 3\right\rangle_i$  of the monoclinic phase were
put on  sites $i\!=\!1,2,3,4$. Red (blue)
indicate positive (negative) lobes.}
\label{orbs}
\end{figure}

At 300$\,$K, LaVO$_{3}$ and YVO$_{3}$ are orthorhombic {paramagnetic} 
Mott insulators.~Their structure   (Fig.$\,$\ref{orbs}) can be derived from
the cubic perovskite ABO$_{3}$, 
with A=La,Y and B=V, by tilting the 
VO$_{6}$ octahedra in alternating directions around the $\mathbf{b}$-axis and
rotating them around the $\mathbf{c}$-axis. 
This GdFeO$_{3}$-type distortion is driven by AO covalency which pulls a given O
atom closer to one of its four nearest A-neighbors  \cite{evad1,njpd1}. 
Since  the Y$\,$4$d$ level is closer to the O$\,$2$p$ level than the La$\,$5$ d $ 
level, the AO covalency increases when going from LaVO$_{3}$ to YVO$_{3}$
and, hence, the shortest AO distance decreases from being 14 to being 20$\,$\% 
shorter than the average, while the angle of tilt increases from $12$ to 
$18^{0},$ and that of rotation  from $7$ to $13^{0}$ \cite{bordet,blake1}. 
Finally, the A-cube is deformed such that one or two of the
ABA body-diagonals is smaller than the average by, respectively, 4 and 10$\,$\% 
in LaVO$_{3}$ and YVO$_{3}$. These  300$\,$K 
structures are determined  
mainly by the strong covalent interactions between O$\,$2$p$ and
the empty B$\,e_{g}$ and A$\,d$ orbitals, hardly by the weak interactions
involving B$\,t_{2g}$ orbitals, and are thus very similar to
the structures of the  $t_{2g}^{1}$ La and Y titanates \cite{evad1,njpd1}.

The $t_{2g}^{2}$ vanadates, however, have a much richer phase diagram than the 
$t_{2g}^{1}$ titanates.
At, respectively, 140$\,$K and $200\,$K, LaVO$_{3}$ and  YVO$_{3}$ transform to a
monoclinic structure in which $\mathbf{c}$ is turned slightly around $%
\mathbf{a}$ whereby the two subcells along \textbf{c}, related by a horizontal
mirror plane in the orthorhombic structure, become
independent (Fig.$\,$\ref{orbs}). 
Most important: 
a sizable (3-4\%)  Jahn-Teller (JT)  elongation of a VO bond,
that along $\mathbf{y}$ in cells 1 and 4, and along 
$\mathbf{x}$ in cells 3 and 2$,$ deforms the VO$_6$ octahedra. 
 At about 140 in LaVO$_3$ and 116$\,$K in YVO$_3$, antiferro{\it magnetic} %
 C-type order develops (FM stacking of AFM $ab$-layers).
At 77$\,$K, YVO$_{3}$ recovers the orthorhombic structure and
the magnetic order changes from C- to G-type (3D-AFM), while
the long VO bond becomes that along $\mathbf{x}$ in cells 
1 and 3, and along $\mathbf{y}$ in 2 and 4. 

It has been suggested \cite{ren2} that these phase transitions 
are driven by the changes in a static orbital order (OO) following 
the observed pattern of JT-distortions \cite{sawada,MF96}. According 
to this JT-OO model, which assumes that
the crystal-field (CF) is due to the oxygen octahedra, the $t_{2g}$
orbital which is \emph{most} antibonding with the O$\,$2$p$ orbitals, i.e. 
$\left\vert sz\right\rangle$ 
where $s$ is the direction of the \emph{short}
in-plane VO bond, is \emph{empty}; the other two $t_{2g}$ orbitals,
due to Hund's rule coupling, are {\em singly} occupied. 
This OO is C-type in the orthorhombic
structure and G-type in the monoclinic structure. Later, this JT-OO
model was challenged by a theory which assumes that the two highest orbitals, 
$\left\vert xz\right\rangle $ and $\left\vert yz\right\rangle,$ are
basically degenerate so that   orbital \emph{fluctuations} play a key role 
\cite{Khalbig,ulrich03}. Recently, ab initio LDA$+U$ \cite{Fang}
calculations gave support to the JT-OO model.

In this Letter we show that  in   LaVO$_3$
quantum effects are strong 
down to 300$\,$K, however they become negligible in the AF-C monoclinic phase.
For YVO$_3$ orbital fluctuations are suppressed already at 
300$\,$K, and the 77$\,$K magnetic transition
is  associated with a change of OO.
We show that the CF splittings result not only from
the JT-,  but also from the GdFeO$_{3}$-type distortions,
and thus OO  is intermediate between C- and G-type. 
The influence of the JT- and the GdFeO$%
_{3}$-type distortion is, respectively, stronger and weaker than in the $%
t_{2g}^{1}$ titanates \cite{mochizuchi,evad1,njpd1}.

The electronic structure is calculated with the LDA+DMFT method \cite%
{lda+dmft}, fully accounting for the orbital degrees of freedom \cite{evad1}. 
First, we compute the LDA bands with the $N^{th}$-order muffin-tin-orbital (NMTO) method \cite{nmto}; 
 we  obtain (for all structures) $\frac{1}{3}$-filled $t_{2g}$ bands separated by a $\sim $0.5$\,$eV gap from the empty $%
e_{g}$ bands and by a $\sim $2$\,$eV gap from the filled O$\,$2$p$ bands.
Next, we L\"{o}wdin downfold to  V$\,t_{2g}$
\begin{table}[tbph]
\centering         
\begin{ruledtabular}
\begin{tabular}{@{}
crcccccrcccccc @{}} $|j\rangle_i$&$\epsilon_j$&& $|xy\rangle$ &$|xz\rangle$&$|yz\rangle$& $n_j$ & $\epsilon_j$&&$
|xy\rangle$ & $|xz\rangle$& $|yz\rangle$&$n_j$ \\ 
\hline
 \multicolumn{2}{c}{ortho}&&\multicolumn{4}{c}{Pbnm LaVO$_3$ (300 K)} &&& \multicolumn{4}{c}{Pbnm YVO$_3$ (300 K)} &\\ $
|1\rangle_1$   & 419    && \phantom{-}.44 & \phantom{-}.24 & \phantom{-}.86 &.78 
                       & 303    && \phantom{-}.56& -.21 &\phantom{-}.80&.96\\ 
$|2\rangle_1$ & 472  && \phantom{-}.34 & \phantom{-}.84 & -.42 & .63 
                       & 383  && \phantom{-}.83&\phantom{-}.17 &-.54&.53\\ 
$|3\rangle_1$ & 511  && -.83 & \phantom{-}.48 & \phantom{-}.29&.59
                       & 510 && -.02 & \phantom{-}.96& \phantom{-}.27&.51 \\   \multicolumn{2}{c}{mono}&&\multicolumn{4
}{c}{P2$_1$/a LaVO$_3$ (10 K)} &&& \multicolumn{4}{c}{P2$_1$/a YVO$_3$ (100 K)}&\\ 
$|1\rangle_1$ &  393   && \phantom{-}.46& \phantom{-}.11& \phantom{-}.88&.82
                       & 285   && \phantom{-}.78 & -.30 & \phantom{-}.55& .97\\ 
$|2\rangle_1$ &471 && \phantom{-}.86& \phantom{-}.16& -.48&.63 
                       &360   &&\phantom{-}.49& -.25& -.83 &.58\\ 
$|3\rangle_1$ &539 && {-}.19& \phantom{-}.98& {-}.03&.55
                       &525 &&\phantom{-}.39 & \phantom{-}.92 & {-}.05 &.55 \\ 
$|1\rangle_3$ &441   && \phantom{-}.71& -.46& {-}.53 &.76 
                       &345     && \phantom{-}.77& \phantom{-}.20 & -.60&.88\\ 
$|2\rangle_3$ &453   && \phantom{-}.08& \phantom{-}.77 & -.64 &.66 
                       &405   && \phantom{-}.62& -.44 & \phantom{-}.65 &.56\\ 
$|3\rangle_3$ &531 && {-}.70 & {-}.41&{-}.58 & .57
                      &547&& {-}.13& {-}.88 & {-}.46 &.45\\ 
\multicolumn{2}{c}{ortho} &&&&&&&&\multicolumn{4}{c}{Pbnm  YVO$_3$ (65 K)}&\\ 
$|1\rangle_1$ &&&&&& &313  &&\phantom{-}.64& -.26 & \phantom{-}.72&.98\\ 
$|2\rangle_1$ &&&&&& &   394 &&\phantom{-}.72 & \phantom{-}.53 & -.45&.60\\ 
$|3\rangle_1 $ &&&&&&& 517 &&-.27 & \phantom{-}.81& \phantom{-}.53&.42\\ 
\end{tabular} \end{ruledtabular} 
\caption{LDA crystal-field (CF) levels wrt the $t_{1g}$ Fermi level, $\protect\epsilon_j $/meV ($j\!=\!1,2,3$), LDA  CF orbitals at site $i$, $\left\vert j\right\rangle _{i}$,
in terms of the cubic orbitals, $|xy\rangle$, 
$|xz\rangle$, and $|yz\rangle$
in the global {\bf x}, {\bf y}, {\bf z} axes  defined in Fig.~\protect\ref{orbs}.
$n_j$ are LDA occupations. 
Orbitals at equivalent sites (see Fig.~\protect\ref{orbs}):
$|j\rangle_2$ ($|j\rangle_4$) is $|j\rangle_1$ ($|j\rangle_3$) with $x\leftrightarrow y$;  for the Pbnm structures  $|j\rangle_3$ is $|j\rangle_1$ with $z\to -z$.
}
\label{cef}
\end{table} 
and remove the energy dependence of the downfolded orbitals by ``$N$-ization" 
\cite{nmto}. These orbitals are strongly localized, having V$%
\,t_{2g}$ character only in their heads. 
Symmetric
orthonormalization finally yields  localized \cite{Lechermann} $t_{2g}$ Wannier functions and their  corresponding Hamiltonian, 
$H^{\mathrm{LDA}}.$ The many-body   Hamiltonian
is then a material-specific $t_{2g}$ Hubbard model, 
$\hat{H}=\hat{H}^{%
\mathrm{LDA}}+\hat{U}$, where for the on-site Coulomb repulsion, $\hat{U},$ we use
the conventional expression \cite%
{fresard}, 
$\hat{U}=\frac{1}{2}\sum\nolimits_{im\sigma,m^{\prime}\sigma^\prime
}U_{m \sigma, m^{\prime }\sigma^\prime}
n_{im\sigma }n_{im^{\prime }\sigma^\prime},$ where
$n_{im\sigma }\!\!=\!\!c_{im\sigma }^{\dagger }c_{im\sigma }^{\phantom{\dagger}}$, and
$c_{im\sigma }^{\dagger }$ creates an electron with spin $\sigma $ in a t$_{2g}$
Wannier orbital $m$  at site $i$. The screened on-site Coulomb interaction is
$U_{m \sigma, m^{\prime }\sigma^\prime}\!=\!U\delta_{m,m^\prime}\delta_{\sigma,-\sigma^\prime} 
+(U^\prime  -J\delta_{\sigma,\sigma^\prime})(1-\delta_{m,m^\prime})$, where
$J$ 
is the exchange term and
$U^{\prime }=U-2J$ the average Coulomb repulsion.
We solve  $\hat{H}$ in  dynamical mean-field theory
 (DMFT)~\cite{DMFT}, using a quantum Monte Carlo \cite{hirsch}
impurity solver and working with the full self-energy matrix, $\Sigma
_{mm^{\prime }}\left( \omega \right) $ \cite{evad1}. Note that inversion is
the only point symmetry of the V sites. The spectral matrix on the real $%
\omega $-axis is obtained by analytic continuation \cite{jarrel}. We use $U%
\mathrm{=}5$~eV and $J\mathrm{=}0.68$~eV, values close to theoretical \cite%
{MF96} and experimental \cite{miyasaka1} estimates, which also
give the correct mass renormalizations/Mott gaps for orthorhombic $%
t_{2g}^{1}$ V/Ti oxides using the same computational scheme \cite%
{evad1,njpd1}.

\begin{figure}[th]
\begin{center}
\rotatebox{0}{\includegraphics[width=0.5\textwidth]{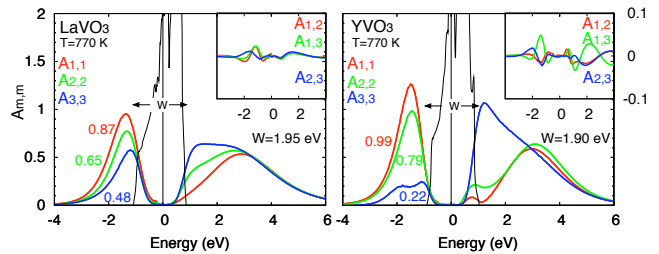}}
\end{center}
\vskip -0.1in
\caption{(color on-line) LDA+DMFT spectral matrix $A_{m,m^\prime}$ in the crystal-field
basis.  
The off-diagonal terms are $ \sim 5$ times smaller than in LDA. 
 In black, the LDA DOS.}
\label{rotdmft}
\end{figure}

Let us start by describing the LDA $t_{2g}$ bands in the
orthorhombic 300$\,$K phase. 
Remarkably, the CF orbitals $\!|j\rangle_i$ (Table~\ref{cef}, $j\!=\!1,\!2,\!3$),
obtained by diagonalizing the on-site $i$  block of $H^{\mathrm{LDA}}$,
the hopping integrals $t_{j,j^\prime}^{i,i^\prime}$ (Table~\ref{tabhopps}), 
the t$_{2g}$ band-shapes  and
band-width  $W$   (Fig.~\ref{rotdmft})
 are
rather similar to those of the $t_{2g}^1$ titanates. 
These  similarities  \cite{compare} are due to  the similarity of  the crystal structures.
Like in the titanates,  the CFs 
are essentially determined by the GdFeO$_{3}$-type distortion, 
mainly via the A-ligand field, specifically the AB and AOB covalency.
However, in the vanadates
the CF splittings are about half those of the
respective titanates and the CF orbitals  \cite{notorb} and the 
hopping integrals 
are less deformed by cation covalency %
\cite{compare}.
This is due to the Ti $\to$ V substitution \cite{notetiv}: 
since V is on the right of Ti in the periodic table,
the  V$\,$3$d$ level is closer to the O$\,$2$p$  and further from the A$\,d$ level
that the Ti$\,$3$d$. Thus the sensitivity of
the B$\,t_{2g}$ Wannier functions to GdFeO$_{3}$-type distortions decreases,
while the sensitivity to JT  increases.

Now, turning on the Coulomb repulsion %
transforms the metallic LDA density of states (DOS) into
the spectral matrix of a Mott insulator (Fig.~\ref{rotdmft}).
 For LaVO$_{3},$ the Mott gap
is $\sim1~$eV, in accord with optical conductivity
data \cite{arima},  and the Hubbard bands are centered
around -1.5~eV and 2.5~eV, in very good agreement with 
photoemission and inverse photoemission \cite{maiti}. For YVO$_{3},$ the gap
is slightly larger, $\sim\!1.2$~eV, in accord with optical data 
\cite{arima}, and the Hubbard bands are centered around
-1.5~eV and 3~eV, in agreement with photoemission \cite{maiti}.

The Mott gaps in the vanadates are somewhat \emph{larger} than in the titanates, for which
the measured gaps are  $\sim\!0.2~$eV in LaTiO$_{3}$ and $\sim\!1\,$eV in YTiO$_{3}$ \cite{arima}, in line with LDA+DMFT results \cite{evad1}. 
This could appear surprising:  
orbital degeneracy increases the critical ratio for the Mott transition, $U_c/W$,
by  {enhancing} the effective band-width, and the enhancement  is 
 stronger the closer the system is to half-filling \cite{Olle}. So
the gap should be smaller for a $t_{2g}^{2}$ than for  a $t_{2g}^{1}$ system, everything else
remaining  the
same. However, the Hund's rule exchange energy, $J$,
strongly suppresses this enhancement, as shown for half-filling
in Ref.~\onlinecite{Han}. For $n\mathrm{=}\frac{1}{6}$ and $n\mathrm{=}%
\frac{1}{3},$ and using a 3-fold degenerate Hubbard model with a rectangular
DOS, $T\mathrm{=}$770$\,$K, and $J/W\mathrm{\sim }\frac{1}{3}$ (like in the
vanadates where $J/W$=0.68/1.9), we find that the metal to insulator transition
occurs for $U^{\prime }/W\sim\,$1.5 when $n\mathrm{=}\frac{1}{6},$ and for $%
U^{\prime }/W\sim\,$1.3 when $n\mathrm{=}\frac{1}{3}.$ So the Hund's-rule
coupling dominates, and thus the vandates can have larger gaps than the titanates.

Like for the titanates \cite{evad1}, 
diagonalization of the DMFT occupation matrix yields
three eigenvectors  nearly identical to the LDA CF
orbitals. 
For LaVO$_{3}$ at 770$\,$K,  the Coulomb repulsion only slightly
increases the orbital polarization by changing the occupations as follows:
0.78$\to$0.87, 0.63$\to$0.65, and 0.59$\to$0.48.
Thus, surprisingly, orbital fluctuations are sizable and remain so  down to room temperature: $n_{3}$=0.26 at 290$\,$K. 
Due to the stronger cation covalency in YVO$_{3}$, the Coulomb repulsion causes substantial orbital polarization already at 770$\,$K
(see Fig.$\,$\ref{FigT}).
At 300 $\,$K, we find that only $ c_{2\sigma }^{\dagger
}c_{1\sigma }^{\dagger }|0\rangle$, paramagnetic  with S=1,
is occupied. 
Thus, YVO$_{3}$ is orbitally ordered well above the magnetic phase transition;
since,  at site 1,  $|3\rangle_1\!\approx \!|xz\rangle$ (see Table~\ref{cef}) is empty   
and thus $|xy\rangle$ and $|yz\rangle$ are $\approx$ singly occupied,  the OO happens to agree with the
prediction of the JT-OO model  \cite{sawada,MF96,Fang}
($|3\rangle_1\approx\!-|3\rangle_3\approx\!|xz\rangle$,
 $|3\rangle_2\approx\!-|3\rangle_4\approx|yz\rangle$), even though the
CF is caused mainly by the GdFeO$_{3}$-type distortion.

What happens in the JT-distorted low temperature phases? The rms values of
the hopping integrals hardly change, so $W$ %
remains 
$\sim1.9$~eV, but individual hopping integrals do change, even in the $%
\left\vert xy\right\rangle ,\left\vert xz\right\rangle ,\left\vert
yz\right\rangle $ representation. Most affected are the CF
orbitals (see Table~\ref{cef}). 

For LaVO$_{3},$ the CF splittings increase; this, in addition to the low temperature,
lets the Coulomb repulsion suppress quantum effects entirely.  
At sites 1 and 2, the occupied state  %
is in accord with the JT 
distortion (at site 1 $|xy\rangle$ and $|yz\rangle$ are singly occupied and $\left\vert 3\right\rangle _{1}\!\approx\! \left\vert
xz\right\rangle$ is empty; at site 2, by symmetry,  the empty state is $\left\vert 3\right\rangle _{2}\!\approx\! \left\vert
yz\right\rangle$; see Fig.~\ref{orbs} and Tab.~\ref{cef})
but this is not the case at sites 3 and 4 
($\left\vert 3\right\rangle _{3}\!\neq\!- \left\vert yz\right\rangle$, $\left\vert 3\right\rangle _{4}\!\neq\! -\left\vert xz\right\rangle$). 
A static mean-field calculation (pseudopotential-based LDA+$U$) \cite{Fang}
yields empty states  not far from ours  \cite{noteempty}, but, without analyzing the results, the
 OO was ascribed to the JT distortions. In contrast, we find
that  the CF orbitals depend crucially also on the GdFeO$_{3}$-type distortions
and that the OO is not of G-type, but is intermediate between C- and G-type.

For YVO$_{3},$ the CF splittings are similar to those of the
300$\,$K phase, but quantum effects are negligible (Fig.$\,$\ref{FigT}).  
On \emph{all} sites in the monoclinic structure
the \emph{empty} orbital is almost the
same as in the orthorhombic 300$\,$K phase so OO does \emph{not} follow the JT distortions ($\left\vert
3\right\rangle _{1}\approx -\left\vert 3\right\rangle _{3}\approx \left\vert
xz\right\rangle $ ,  $\left\vert
3\right\rangle _{2}\approx -\left\vert 3\right\rangle _{4}\approx \left\vert
yz\right\rangle $), 
but is almost C-type.  
In the orthorhombic 77$\,$K phase,  
the empty orbital at site 1, $\left\vert 3\right\rangle_1,$ only
roughly equals $\left\vert xz\right\rangle .$ 
Our results are consistent with resonant x-ray
scattering \cite{Noguchi} and magnetization \cite{ren2} data. 
LDA+$U$  \cite{Fang} yields results close to ours  \cite{noteempty}.

\begingroup
 \begin{table*}[htbp]
   \centering
   \begin{ruledtabular}
   \begin{tabular}{@{}ccrrrrrrrrrrrrrrrrrrrrrr@{}} 
    &&\multicolumn{3}{c}{La Pbnm\; } 
       &\multicolumn{3}{c}{La P2$_1$/a, site 1\:   } 
      &\multicolumn{3}{c}{La P2$_1$/a, site 3\;}
      &\multicolumn{3}{c}{Y Pbnm\;}
       &\multicolumn{3}{c}{Y P2$_1$/a, site 1\; }
        & \multicolumn{3}{c}{Y P2$_1$/a, site 3\;}
       &\multicolumn{4}{c}{Y Pbnm (65 K)}\\
       {${j,j^\prime}\backslash{lmn}$} && 001 & 100&010 \;& 001&100&010\:&001&100&010   
 \;&001 &100&010 \;&001&100&010\:&001&100&010 \;&001&100&010\;\\
 \cline{3-5\!} \cline{6-8\;} \cline{9-11\;} \cline{12-14\;} \cline{15-17\;} \cline {18-20\;} \cline {21-23\;}
$1,1$  &&130    &-65   &-65   \;        
             & 85    &-39&-39\: &85&-159&-159 \; &             
            -13 &-17&-17 \; 
            & -49 &-84&-84\:  &-49& -92&-92    \;                          
               & -35 &-34&-34 \; %
                      \\
$1,2$   &&  9 &-37&-198 \; 
            &27 &-110&-127\:  &-36&-65&98 \; &
           -63 &-102&-157 \; 
             &-20 &-117&-62\:  &-46&-73&-169 \;
              &  -38 &-66&-195  \;                   
\\
$1,3$    &&119 &104&-7 \; 
             &154 &31 &-155\:&153&90&23 \; &
                   \phantom{-1}46 &66&-138 \; 
                     &30 &11&-170\:  &\phantom{-1}26&\phantom{-1}80&-91 \;
                     & 52   &\phantom{-}100 &-68 \;
                    \\
$2,2$ &&193 &47&47 \;
         &-133 &-84&-84\:&-133&94&94 \;&
          86&-48&-48 \;
          &72&-6&-6\: &72&25&25 \;
          &142&-28&-28 \; 
\\
$ 2,3$ && 26&13&9 \;&
           -57 &76&73\: &-140&110&30 \;&
          38 &5&20 \;
          &-112 &9&94\: &118&-41&30 \; 
          &67&-27&7 \; \\
$3,3$ && 36 &-152&-152 \;&
65& -38 &-38\:  &65& -109&-109 \;&
202&-66&-66 \;&
183 &-48 & -48\: &  183&-63&-63 \;&
173& -61&-61 \;
          \end{tabular} \end{ruledtabular}
          
  \caption{Hopping integrals   
               $t_{j,j^\prime}^{i, i^\prime}/$meV 
                from site $i$  to a site  $i^\prime=i+l\mathbf{x}+m\mathbf{y}+n\mathbf{z}$,
                in the basis ($j,j^\prime$) of crystal-field orbitals. Here
                 $i\!=\!1$ and (P2$_1$/a only) $i\!=\!3$. 
            Notice that $t_{j,j^\prime}^{i , i+\mathbf{z} }=t_{j^\prime,j}^{i+\mathbf{z} , i}$ and 
                    $t_{ j, j^\prime}^{i, i+\mathbf{x}}=t_{j^\prime,j}^{i , i+\mathbf{y}}$. 
                                        For Pbnm structures only: $t_{j,j^\prime}^{i, i+\mathbf{z}}=
                                        t_{j^\prime,j}^{i\; i+\mathbf{z}}$. }
    \label{tabhopps}
\end{table*}
\endgroup
Finally, we find that the monoclinic structure
favors C-type magnetic order over G-type by increasing some hopping
integrals $t^{i,i^\prime}_{j,j^\prime}$ (Table~\ref{tabhopps}) to the empty orbital $|3\rangle $ along the $\mathbf{c}$%
-direction. 
Assuming complete OO, conventional theory
yields, for the superexchange couplings, 
$$J_{\mathrm{SE}}^{i,i^\prime} \sim\! 
\frac{1+J/U}{U+2J} 
\sum
_{j,j^{\prime} \le 2}
{{|t_{jj^{\prime}}^{i, i^\prime}|^{2}}} -
\frac{J/U}{U-3J}
\sum
_{j\le2} \left(
{|t_{j3}^{i, i^\prime}|^{2}\!+\!|t_{3j}^{i, i^\prime}|^{2} } \right)
$$
with  $j,j^\prime$  CF orbitals and  $i,i^\prime$ neighboring sites.  
 We find that C-type order 
($J_{\mathrm SE}^{i,i+\mathbf{z} }\!<\!0$, $J_{\mathrm SE}^{i,i+\mathbf{x}}\!=J_{\mathrm SE}^{i,i+\mathbf{y}}\!>\!0$)
is favored over G-type, for which all couplings are positive, 
if $J/U \gtrapprox 0.16$.
While the actual values of $J^{i , i^\prime}_{\mathrm{SE}}$ are sensitive to details \cite{njpd1,solovyev06,Fang},
this 
provides a microscopic explanation of  C-type 
 order in  monoclinic LaVO$_{3}$ and YVO$_{3}$,
the change from C- to G-type across the structural phase transition in YVO$%
_{3}$, and thus could also explain the magnetization-reversal  phenomena \cite%
{ren2}.

In conclusion, we find that the orthorhombic  LaVO$_3$  is one
of the few Mott insulators which exhibits large quantum effects at room
temperature.
This is not the case  for YVO$_{3}$ (and $t_{2g}^1$ titanates \cite{evad1}).
In the low temperature phases, 
orbital fluctuations are negligible for both vanadates.  
 This  supports the view \cite{ren2,MF96,sawada} 
that  the magnetic structures  of the vanadates can be explained by 
orbital-order.
Recent LDA+$U$ \cite{Fang} and LDA+PIRG \cite{Imada} calculations
agree with this, but previous literature ascribed OO mainly to JT-distortions.
In contrast, we proved  that both
the JT {\em and} the GdFeO$_{3}$-type distortions are crucial for the
CF orbitals and their hopping integrals, and thus for the type of orbital
and magnetic order.
The effects of the GdFeO-type distortions are weaker 
and those of JT stronger than in $t_{2g}^1$ titanates; 
their interplay is responsible for the rich phase diagram of the vanadates.

We thank  E.~Koch, A.I.~Lichtenstein, S.~Biermann, and A.~Georges
for discussions and J.~Nuss for graphics support.
Computations were done on the J\"{u}lich BlueGene.
M.~D. thanks the MPG Partnergroup  program.


\begin{thebibliography}{99}
\bibitem{notetiv} To prove this
we performed LDA calculations using the
titanates crystal structure and replacing Ti with V.
\bibitem{noteempty}In Ref.~\cite{Fang}
occupations are not  for Wannier functions ($n_{t_{2g}}<2$),
so a full comparison is not possible.
\end{thebibliography}
\end{document}